 \documentclass[pmlr,twocolumn,10pt]{jmlr} 

\usepackage{multicol}
\usepackage{booktabs}
\usepackage{siunitx}

\newcommand{\equal}[1]{{\hypersetup{linkcolor=black}\thanks{#1}}}
\newcommand{\indtable}[1]{\hspace{1.2em}#1}

\theorembodyfont{\upshape}
\theoremheaderfont{\scshape}
\theorempostheader{:}
\theoremsep{\newline}

\jmlrvolume{LEAVE UNSET}
\jmlryear{2023}
\jmlrsubmitted{LEAVE UNSET}
\jmlrpublished{LEAVE UNSET}
\jmlrworkshop{Machine Learning for Health (ML4H) 2023} 

\title[Multi-modal Transformer for Breast Cancer Classification and Risk Assessment]{Leveraging Transformers to Improve Breast Cancer Classification and Risk Assessment with Multi-modal and Longitudinal Data}

 \author{\vspace{-50pt}}

\author{%
\Name{Yiqiu Shen}$^{1,2,3}$\equal{These authors contributed equally} \Email{ys1001@nyu.edu}\\
\Name{Jungkyu Park}$^3$\footnotemark[1] \Email{jp.park@nyu.edu}\\
\Name{Frank Yeung}$^3$ \Email{frank.yeung@nyulangone.org}\\
\Name{Eliana Goldberg}$^2$ \Email{Eliana.Goldberg@nyulangone.org }\\
\Name{Laura Heacock}$^{2,3}$ \Email{Laura.Heacock@nyulangone.org }\\
\Name{Farah Shamout}$^{4,5,6}$ \Email{fs999@nyu.edu}\\
\Name{Krzysztof J. Geras}$^{2,3,1}$\Email{k.j.geras@nyu.edu}\\
\addr$^1$Center for Data Science, New York University, New York, NY, USA\\
\addr$^2$Department of Radiology, NYU Langone Health, New York, NY, USA\\
\addr$^3$NYU Grossman School of Medicine, New York, USA\\
\addr$^4$Division of Engineering, New York University Abu Dhabi, Abu Dhabi, UAE\\
\addr$^5$Computer Science and Engineering, New York University, New York, USA\\
\addr$^6$Department of Biomedical Engineering, New York University, New York, USA\\
}

\begin{document}
\maketitle
\begin{abstract}
Breast cancer screening, primarily conducted through mammography, is often supplemented with ultrasound for women with dense breast tissue. However, existing deep learning models analyze each modality independently, missing opportunities to integrate information across imaging modalities and time. In this study, we present Multi-modal Transformer (MMT), a neural network that utilizes mammography and ultrasound synergistically, to identify patients who currently have cancer and estimate the risk of future cancer for patients who are currently cancer-free. MMT aggregates multi-modal data through self-attention and tracks temporal tissue changes by comparing current exams to prior imaging. Trained on 1.3 million exams, MMT achieves an AUROC of 0.943 in detecting existing cancers, surpassing strong uni-modal baselines. For 5-year risk prediction, MMT attains an AUROC of 0.826, outperforming prior mammography-based risk models. Our research highlights the value of multi-modal and longitudinal imaging in cancer diagnosis and risk stratification.
\end{abstract}

\begin{keywords}
breast cancer, deep learning, mammography, ultrasound, multi-modal data
\end{keywords}

\vspace{-12pt}
\section{Introduction}
\label{sec:intro}
Breast cancer is the leading cause of cancer death in women globally. Breast cancer screening aims to detect cancer in its early stage of development so that treatment can lead to better patient outcomes. Despite the wide adoption of full-field digital mammography (FFDM) and digital breast tomosynthesis (DBT), only approximately 75\% of breast cancers are diagnosed through mammography~\citep{lee2021radiologist, monticciolo2017breast}. This limitation stems from dense breast tissue obscuring smaller tumors and reducing mammography's sensitivity to as low as 61-65\% in women with extremely dense breasts~\citep{mandelson2000breast, wanders2017volumetric, destounis2017using}. These women require supplemental screening to compensate for the limitations of mammography. Ultrasound is commonly used given its accessibility, lower costs, and lack of radiation. While ultrasound does increase cancer detection rates by 3-4 per 1000 women~\citep{berg2012detection}, this improvement comes at the cost of lower specificity, increased recall rates of 7.5-10.6\%~\citep{berg2019screening,brem2015screening,butler2020screening} and lower positive predictive values (PPV) of 9-11\%~\citep{berg2019screening}, leading to unnecessary diagnostic imaging and biopsies. Artificial intelligence presents opportunities to improve precision by synergistically using mammography and ultrasound.

Deep learning models have been applied to support breast cancer screening, primarily through detecting existing cancers~\citep{shen2021interpretable,wu2019deep,mckinney2020international,shen2019deep,lotter2021robust,rodriguez2019stand,lotter2021robust} or predicting future risk~\citep{yala2019deep, yala2021toward, arasu2023comparison,lehman2022deep}. Within this area, several seminal studies have made great contributions. ~\cite{mckinney2020international} demonstrated that convolutional neural networks (CNNs) match the screening performance of radiologists and retain generalizability across countries. ~\cite{yala2019deep, yala2021toward} proposed Mirai, an AI system that utilizes mammography and clinical risk factors to forecast the future risk of breast cancer. ~\cite{shen2021artificial} showed that AI can reduce the false-positive rates by 37.3\% in breast ultrasound interpretation, without compromising sensitivity. 

Despite these advances, there are two major limitations. First, existing work tends to concentrate on a single imaging modality, ignoring cross-modal patterns only noticeable through integrating multiple imaging modalities. In contrast, radiologists often use complementary imaging modalities to ascertain a diagnosis and increase accuracy~\citep{bankman2008handbook}. Furthermore, existing work overlooks the utility of prior imaging despite that comparison to prior mammograms has been shown to significantly reduce the recall rate and increase cancer detection rate and PPV1~\citep{hayward2016improving}.

In this study, we introduce an AI system capable of referencing prior imaging and synthesizing information from both mammography and ultrasound. This system has two functionalities: detecting extant cancers and predicting future cancer risk. 

\begin{figure}[ht!]
    \centering
\includegraphics[width=0.4\textwidth]{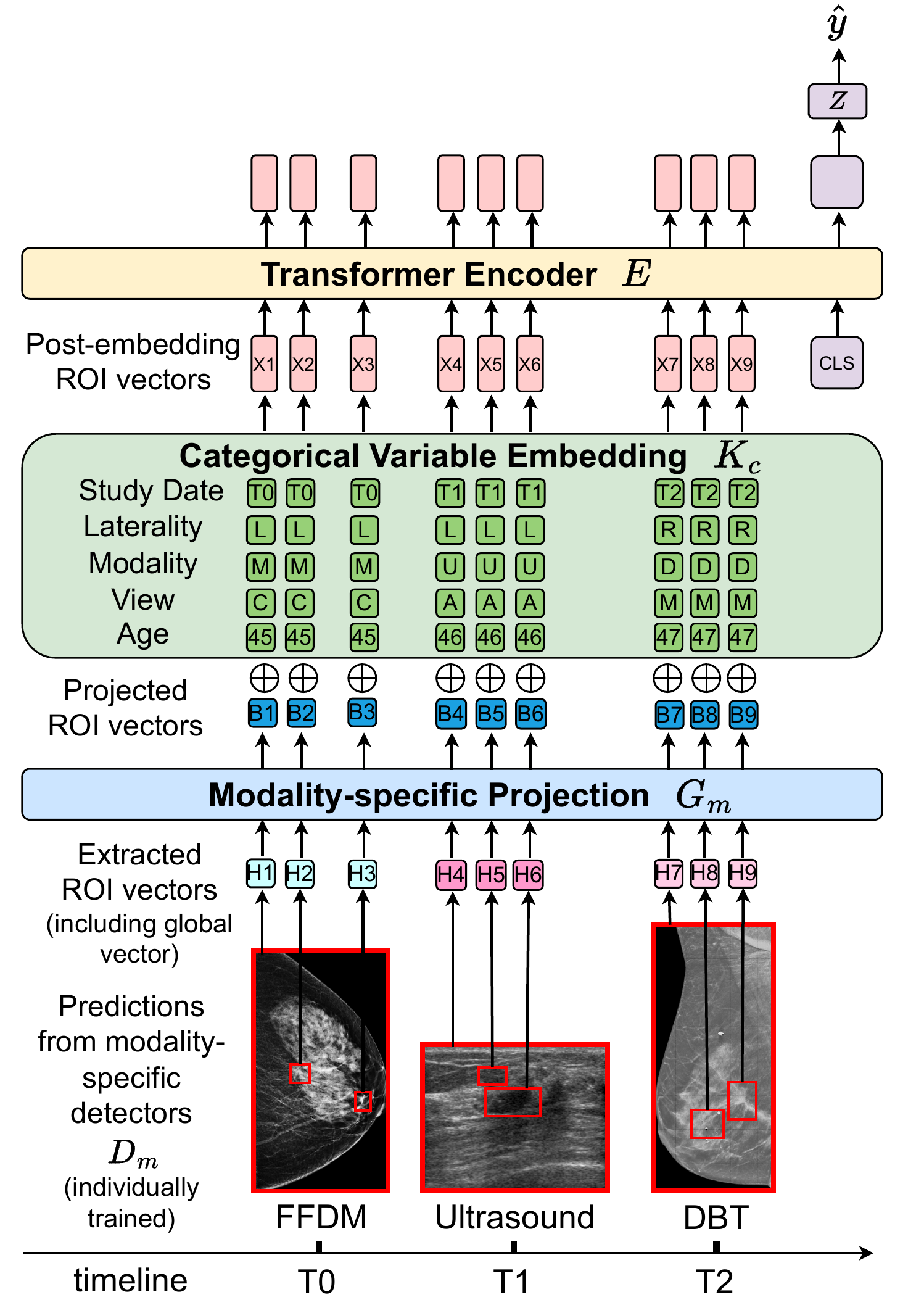}
    \caption{Architecture of MMT.}
    \label{fig:mml_transformer}
\end{figure}

\vspace{-12pt}
\section{Methods}
\label{sec:method}
\paragraph{Problem Formulation}
We formulate breast cancer diagnosis as a sequence classification task. Let $S_i$ denote an imaging exam with images $I_i^1, I_i^2, ..., I_i^{l_i}$, where $l_i$ denotes the number of images in $S_i$. Let $m_i$ denote $S_i$'s imaging modality and $t_i$ denote the time when $S_i$ is performed. $S_i$ has prior exams $S_i^1, S_i^2, \ldots, S_i^{r_i}$, all of which belong to the same patient but were performed at the same or earlier times $t_i^1, t_i^2, ..., t_i^{r_i} \leq t_i$, where $r_i$ is the number of prior exams. Our goal is to build an AI system that takes the sequence $Q_i = \{S_i, S_i^1, S_i^2, ..., S_i^{r_i}\}$ as an input and makes a series of probabilistic predictions $\hat{y}_i^j \in [0,1]$ quantifying the probability of malignancy within 120 days ($j=0$) and each of 1 to 5 years ($j=\text{1, 2, ..., 5}$) from $t_i$.

There are two key challenges. First, each patient has a unique exam history (hence a unique set of imaging modalities) with variable numbers of images. The model must handle this variability. Second, malignant lesions have diverse visual patterns across modalities. The model must capture this spectrum and integrate findings across imaging modalities.

\paragraph{Multi-modal Transformer} We propose the Multi-modal Transformer (MMT) to address the aforementioned challenges. As illustrated in Figure~\ref{fig:mml_transformer}, MMT produces cancer predictions in three steps. First, MMT applies modality-specific detectors to all images to extract feature vectors from regions that are suspicious of cancer. Second, it combines these features with embeddings of non-image variables. Finally, the post-embedding features are fed into a transformer encoder to detect temporal changes in tissue patterns, integrate multi-modal tissue information, and produce malignancy predictions. The following paragraphs elaborate on each step in detail.

\textit{Generating regions of interest and feature vectors.} A typical input sequence $Q_i$ contains images of multiple modalities. Since tumor morphology varies across modalities, we train a detector $D_{m}$ for each modality $m \in \{\text{FFDM},\text{DBT}, \text{Ultrasound}\}$. Each $D_m$ takes an image $I$ as input and outputs $k_m$ regions of interest (ROIs) with feature representations $H \in \mathbb{R}^{k_m, d_m}$ and scores $P_c \in \mathbb{R}^{k_m}$, reflecting its belief that each ROI contains a malignant lesion. Only the top $k_m$ highest scoring ROIs are extracted, where $k_m$ is a hyper-parameter that is tuned on the validation set. Some detectors also produce global feature vectors. As feature vectors are extracted by different $D_{m}$, they vary in size and scale. To address this, we apply a modality-specific transform $G_{m}$ (a multilayer perceptron) to project all features into the same  space: $B = G_m(H)$, 
where $B \in \mathbb{R}^{k_m, d}$ denotes the post-projection feature vectors.

\textit{Categorical embeddings.} We incorporate categorical variables including study date, laterality, imaging modality, imaging view angles, and patient age discretized into ranges. These variables help our model understand the temporal and spatial patterns in the ROI patches. We utilize the embedding technique to map each variable $c$ to an embedding vector $K_c \in \mathbb{R}^{100}$, which are then concatenated with the post-projection ROI feature vectors. Next, we use a multilayer perceptron $f_\text{emb}: \mathbb{R}^{d+500} \mapsto \mathbb{R}^{d}$ to reduce dimensionality:
\begin{equation}
    X = f_\text{emb}([B, K_{\text{date}}, K_{\text{lat}}, K_{\text{mod}}, K_{\text{view}}, K_{\text{age}}]^T),
\end{equation}
where $X \in \mathbb{R}^{k_m, d}$ denote post-embedding vectors.

\textit{Transformer.} We use a transformer encoder to enable interaction among post-embedding ROI vectors from all images~\citep{vaswani2017attention}. The transformer encoder uses multi-head attention to selectively combine information from the input sequence. As a common practice~\citep{devlin2018bert}, we inject a special CLS token into the input sequence. The CLS token condenses variable-length input sequences into a fixed-size aggregated representation and allows the transformer to iteratively update it using signals from all post-embedding ROI vectors:
\begin{equation}
    \text{CLS'} = \text{transformer}([X, \text{CLS}]).
\end{equation}

Next, we apply a multi-layer perceptron $Z: \mathbb{R}^d \mapsto \mathbb{R}^6$ with a rectified linear unit (ReLU) on the post-transformer CLS vector (CLS') to generate six non-negative risk scores for non-overlapping intervals: baseline risk within 120 days ($L^0$), additional 120d-1yr risk ($L^1$), 1-2yr risk ($L^2$), 2-3yr risk ($L^3$), 3-4yr risk ($L^4$), and 4-5yr risk ($L^5$). This is expressed in the equation below:
\begin{equation}
    [L^0, L^1, L^2, L^3, L^4, L^5]^T = \text{ReLU}(Z(\text{CLS'})).
\end{equation}

Finally, we use an additive hazard layer~\citep{aalen2005aalen} with sigmoid non-linearity to generate the cumulative probability of malignancy:
\begin{equation}
    \hat{y}^j = \sigma(L^0 + \sum_{k=1}^j L_k).
\end{equation}

\paragraph{Training}
MMT is trained in two phases. First, for each modality, we independently train cancer detectors using only images from that modality. FFDM and DBT detectors are parameterized as YOLOX~\citep{ge2021yolox}, MogaNet~\citep{li2022efficient}, and GMIC~\citep{shen2021artificial,shen2019globally}. YOLOX is an anchor-free version of YOLO~\citep{redmon2016you}, a popular object detection model family. MogaNet is a CNN that can efficiently model interactions among visual features. GMIC is a resource-efficient CNN that is designed for high-resolution medical images. For DBT, we train on 2D slices to limit computation. YOLOX and MogaNet detectors are trained on both image and bounding box labels. To train with image labels, we attention-pool the features of the highest-scoring boxes and classify them using a logistic regression layer. We use the UltraNet proposed in~\cite{shen2021artificial} as our ultrasound detector. To balance computation cost, we extracted 10 ROIs from each image across modalities.

In the second phase, we freeze detectors and train the transformer encoder, MLPs, and embeddings on multi-modal sequences using binary cross-entropy loss and Adam optimizer~\citep{kingma2014adam} with a learning rate set to $10^{-5}$.

\paragraph{Ensembling} To improve results, we use model ensembling~\citep{dietterich2000ensemble}. We train 100 MMT models, randomly combining one model from each detector family with a transformer encoder that is randomly parameterized as either a DeiT~\citep{touvron2021training}, a ViT~\citep{dosovitskiy2020image}, or a BERT~\citep{devlin2018bert}. The top 5 MMT models by validation performance are ensemble averaged to produce the final prediction.

\vspace{-12pt}
\section{Results}
\label{sec:result}
\paragraph{Dataset} We trained and evaluated our model on the NYU Breast Cancer Diagnosis Multi-modal Dataset containing 1,353,521 FFDM/DBT/ultrasound exams from 297,751 patients who visited NYU Langone Health between 2010 and 2020. Exams are split into training (87.1\%), validation (3.9\%), and test (8.9\%) sets, with each patient's exams assigned to only one set. Labels indicating presence or absence of cancer are derived from corresponding pathology reports. Validation and test sets are filtered so cancer-positive exams have pathology confirmation and cancer-negative exams have a negative follow-up. See Table~\ref{tab:dataset_table} for dataset details. 

\paragraph{Evaluation} We evaluate our model's ability to detect existing cancers and predict risk in the general screening population. Each test case is a screening visit with required FFDM and optional DBT/ultrasound. Our model utilizes all available modalities and prior studies. For detection of existing cancer, a visit is cancer-positive if a pathology study within 120 days of imaging confirms cancer. The 121,037 exams in the test set came from 54,789 screening visits, out of which 483 led to cancer diagnosis. Among these visits, 26,028 (47.5\%) had same-day DBT, 15,542 (28.4\%) had same-day ultrasound, and 34,888 (63.7\%) had prior imaging studies available. On average, each visit had 1.9 associated prior imaging exams. For 5-year risk stratification, we exclude screening-detected cancers and negative cases with $<$ 5-year follow-up, focusing solely on long-term prediction. This leaves 6,173 visits with 598 positive cases in the test set. We use area under the ROC curve (AUROC) and area under the precision-recall curve (AUPRC) as evaluation metrics.

\paragraph{Performance} For cancer diagnosis, we compared MMT to four baselines – GMIC, YOLOX, and MogaNet using FFDM only, and a multi-modal ensemble averaging predictions from the mammogram baselines and UltraNet, when ultrasound is available. For each uni-modal baseline, we trained 20 models and ensembled the top 5 for evaluation. We report the performance in Table~\ref{tab:perf-cancer-120}. MMT achieved higher AUROC and AUPRC than mammogram-only baselines, indicating that incorporating ultrasound improves diagnostic accuracy. MMT also outperformed the multi-modal ensemble, indicating the transformer integrates multi-modal information better than simple averaging. 

\begin{table}
\caption{Cancer diagnosis performance.}
\label{tab:perf-cancer-120}
\centering
\begin{tabular}{lcccc}
\toprule
Model   & AUROC & AUPRC  \\
\bottomrule
GMIC~\citep{shen2021interpretable} & 0.866 & 0.167 \\
YOLOX~\citep{ge2021yolox} & 0.876 & 0.172 \\
MogaNet~\citep{li2022efficient} & 0.874 & 0.181 \\ \midrule
Multi-modal Ensemble & 0.925 & 0.251	 \\
MMT & \textbf{0.943} & \textbf{0.518}\\
\bottomrule
\end{tabular}
\end{table}

For risk stratification, we compared MMT to two baselines: radiologists' BI-RADS diagnosis and Mirai~\citep{yala2021toward}, an AI system predicting future breast cancer risk using both categorical risk factors and mammograms, on the same test set. We report the performance in Table~\ref{tab:perf-risk}. For 5-year cancer prediction, MMT achieved an AUROC of 0.826 and AUPRC of 0.524, outperforming both methods. By leveraging multi-modal imaging and longitudinal patient history, MMT demonstrates a strong ability to predict future breast cancer risk.

\vspace{-10pt}
\begin{table}[ht]
\caption{Risk stratification performance.}
\label{tab:perf-risk}
\centering
\begin{tabular}{lcc}
\toprule
Model & AUROC & AUPRC\\
\bottomrule
BI-RADS & 0.585 & 0.118\\
Mirai~\citep{yala2019deep} & 0.732 & 0.252\\
MMT & \textbf{0.826} &  \textbf{0.524}\\
\bottomrule
\end{tabular}
\end{table}

\paragraph{Ablation Study} We performed an ablation study to understand the impact of supplemental modalities and prior imaging. We evaluated MMT using only mammograms with prior imaging (mammo only), using both mammograms and ultrasound but with no prior imaging (no prior), and incorporating only 1, 2, or 3 years of prior imaging. As shown in Table~\ref{tab:ablation}, compared to mammograms alone, MMT shows meaningful performance gains when ultrasound is also used for both tasks, confirming the importance of supplemental modality. In contrast, prior imaging mainly contributes to long-term risk stratification. Moreover, prior imaging beyond two years provides only marginal improvement. This observation is consistent with the clinical practice of using up to two years as references. Overall, the ablation highlights the value of multi-modal and longitudinal information, with ultrasound and recent prior imaging improving cancer diagnosis and risk prediction.

\begin{table}
\caption{Ablation study on the impact of supplemental modality and prior imaging.}
\label{tab:ablation}
\centering

\resizebox{.49\textwidth}{!}{
\begin{tabular}{lcccc}
\toprule
& \multicolumn{2}{c}{Cancer Diagnosis} & \multicolumn{2}{c}{Risk Stratification}\\
Model & AUROC & AUPRC & AUROC & AUPRC\\
\bottomrule
mammo only & 0.925 & 0.373 & 0.814 & 0.516 \\ \midrule
no prior & 0.944 & 0.484 & 0.817 & 0.513\\
1 year prior & 0.944 & 0.487 & 0.816 & 0.514 \\
2 year prior & \textbf{0.945} & 0.507 & 0.822 & 0.521 \\
3 year prior & 0.944 & 0.512 & 0.825 & 0.523\\ \midrule 
MMT & 0.943 & \textbf{0.518} & \textbf{0.826} & \textbf{0.524}\\
\bottomrule
\end{tabular}
}
\vspace{-12pt}
\end{table}
\vspace{-8pt}

\section{Discussion and conclusion}
\label{sec:discussion}
In medical imaging, each modality has its own strengths and limitations. Radiologists often combine multiple modalities to inform decision-making. In this spirit, we propose MMT to jointly utilize mammography and ultrasound for breast cancer screening. MMT achieves strong performance in identifying existing cancers and predicting long-term risk. 

Standard-of-care risk models use personal history, genetics, family history, and breast density to estimate risk, but exhibit suboptimal accuracy~\citep{arasu2023comparison}. This stems from their reliance on coarse variables that inadequately capture underlying breast tissue heterogeneity associated with cancer risk. Our study demonstrates that integrating multi-modal longitudinal patient data with neural networks can significantly improve risk modeling by extracting richer tissue features predictive of cancer development.

This study also has limitations. First, we did not evaluate our model on external datasets, which is necessary to demonstrate its generalizability across diverse patient populations and acquisition protocols. Additionally, our model uses primarily imaging data. Incorporating risk factors like demographics and family history could further enhance risk modeling. Addressing these limitations through multi-institutional studies, and integrating non-imaging data will be important future directions to translate these promising results into clinical practice.

\section{Acknowledgements}
This research was supported by the National Institutes of Health (NIH), through grants TL1TR001447 (the National Center for Advancing Translational Sciences) and P41EB017183, the Gordon and Betty Moore Foundation (9683), and the National Science Foundation (1922658). The content is solely the responsibility of the authors and does not necessarily represent the official views of any of the bodies funding this work. 


\setcounter{table}{0}
\renewcommand{\thetable}{A\arabic{table}}

\begin{table*}[ht]
\caption{Performance of MMT on patient subgroups stratified by age and breast density.}
\label{tab:subgroup}
\centering
\begin{tabular}{lcc} \toprule
 & AUROC (cancer diagnosis) & AUROC (risk stratification)\\ \hline
\textbf{Age} & & \\ \hline
$<$50 years old & 0.927 & 0.820  \\ 
50-60 years old & 0.957 & 0.790 \\ 
60-70 years old & 0.941 & 0.810 \\ 
70-80 years old & 0.920 & 0.830  \\ 
$>$80 years old & 0.928 & 0.965 \\ \hline
\textbf{Breast Density}  \\  \hline
A & 0.967 & 0.776  \\ 
B & 0.949 & 0.813  \\ 
C & 0.936 & 0.824 \\ 
D & 0.921 & 0.838 \\ \hline

\end{tabular}
\end{table*}

\begin{table*}[ht!]
\caption{Characteristics of the NYU Breast Cancer Diagnosis Multi-modal Dataset.}
\label{tab:dataset_table}
\centering
\small
\begin{tabular}{lccc} \toprule
 & Training & Validation & Test \\ \hline
\textbf{Patients} & 263,573 & 10,839 & 23,339 \\ \hline
Age, mean years (SD) & 56.98 (13.09) & 59.58 (11.55) & 59.77 (11.47) \\ 
\indtable $<$40 yr old& 17,186 (6.52\%) & 173 (1.60\%) & 351 (1.50\%) \\ 
\indtable 40-49 yr old& 63,158 (23.96\%) & 2,308 (21.29\%) & 4,806 (20.59\%) \\ 
\indtable 50-59 yr old& 64,882 (24.62\%) & 3,088 (28.49\%) & 6,679 (28.62\%) \\ 
\indtable 60-69 yr old& 54,980 (20.86\%) & 2,960 (27.31\%) & 6,582 (28.20\%) \\ 
\indtable $\geq70$ yr old& 43,022 (16.32\%) & 2,092 (19.30\%) & 4,582 (19.63\%) \\ 
\indtable Unkown & 20,345 (7.72\%) & 218 (2.01\%) & 339 (1.45\%) \\ \hline 
Breast density & & \\ 
\indtable A & 18,616 (7.06\%) & 773 (7.13\%) & 1,756 (7.52\%) \\ 
\indtable B & 93,990 (35.66\%) & 4,225 (38.98\%) & 9,415 (40.34\%) \\ 
\indtable C & 113,631 (43.11\%) & 5,005 (46.18\%) & 10,542 (45.17\%) \\ 
\indtable D & 16,991 (6.45\%) & 618 (5.70\%) & 1,287 (5.51\%) \\ 
\indtable Unkown & 20,345 (7.72\%) & 218 (2.01\%) & 339 (1.45\%) \\ \hline 
\textbf{Exams} & 1,179,171 & 53,313 & 121,037 \\ \hline 
\indtable cancer 120 days & 15,586 (1.32\%) & 877 (1.65\%) & 1,902 (1.57\%) \\ 
\indtable cancer 120 days - 1 year & 1,566 (0.13\%) & 19 (0.04\%) & 13 (0.01\%) \\ 
\indtable cancer 1-2 years & 4,816 (0.41\%) & 54 (0.10\%) & 133 (0.11\%) \\ 
\indtable cancer 2-3 years & 3,417 (0.29\%) & 249 (0.47\%) & 613 (0.51\%)\\ 
\indtable cancer 3-4 years & 2,089 (0.18\%) & 182 (0.34\%) & 364 (0.30\%) \\ 
\indtable cancer 4-5 years & 1,281 (0.11\%) & 101 (0.19\%) & 248 (0.20\%) \\ 
\indtable cancer $>$5 years & 1,391 (0.12\%) & 92 (0.17\%) & 268 (0.22\%) \\ \hline 
Imaging modality \\ 
\indtable FFDM & 546,862 (46.38\%) & 26,687 (50.06\%) & 62,249 (51.43\%) \\ 
\indtable DBT & 300,277 (25.47\%) & 12,139 (22.77\%) & 28,339 (23.41\%) \\ 
\indtable Ultrasound & 332,032 (28.16\%) & 14,487 (27.17\%) & 30,449 (25.16\%) \\ \hline 
Exam-level BI-RADS\\ 
\indtable 0 & 439,112 (10.99\%) & 19,811 (9.66\%) & 45,060 (9.57\%) \\ 
\indtable 1 & 439,112 (37.24\%) & 19,811 (37.16\%) & 45,060 (37.23\%) \\ 
\indtable 2 & 501,915 (42.57\%) & 24,598 (46.14\%) & 56,810 (46.94\%) \\ 
\indtable 3 & 70,935 (6.02\%) & 1,683 (3.16\%) & 3,453 (2.85\%) \\ 
\indtable 4 & 33,517 (2.84\%) & 1,999 (3.75\%) & 4,002 (3.31\%) \\ 
\indtable 5 & 2,535 (0.21\%) & 73 (0.14\%) & 129 (0.11\%) \\ 
\indtable 6 & 1,611 (0.14\%) & 0 (0.00\%) & 0 (0.00\%) \\ \hline 

\end{tabular}
\end{table*}

\end{document}